\documentclass[conference]{IEEEtran}

\ifCLASSINFOpdf
  
\else
 
\fi

\usepackage{amsmath,graphicx,amsthm,amssymb}
\usepackage{graphicx}
\usepackage{epsfig,epstopdf}
\usepackage{booktabs}
\usepackage{multirow}
\usepackage{chngpage}
\usepackage{algorithm}
\usepackage{longtable}
\usepackage{algorithmic,color}
\usepackage{subfigure}

\usepackage{float}
\usepackage{multicol} 
\usepackage{url}
\usepackage{fancyhdr}
\usepackage{balance}

\makeatletter
\def\endthebibliography{%
	\def\@noitemerr{\@latex@warning{Empty 'thebibliography' environment}}%
	\endlist
}
\makeatother

\begin{document}

\title{Generative Adversarial Networks for Real-time Stability of Inverter-based Systems}

\author{\IEEEauthorblockN{Xilei Cao, Gurupraanesh Raman, Gururaghav Raman, and Jimmy Chih-Hsien Peng}
\IEEEauthorblockA{Department of Electrical and Computer Engineering\\
National University of Singapore, Singapore 117583\\
Email: jpeng@nus.edu.sg}
}

\maketitle

\begin{abstract}

In islanded systems with droop-controlled sources, the droop coefficients need to be tuned in real-time using supervisory control to maintain asymptotic stability. In contrast to offline tuning methods, online domain-of-stability estimation yields non-conservative droop gains in real-time, ensuring good power sharing performance as the operating point varies. The challenge in the conventional online domain-of-stability estimation process is its unscalability and high computational complexity. In this paper, an efficient alternative using conditional Generative Adversarial Networks (cGANs) is described. We demonstrate that the notion of power system stability can be learned by such deep neural networks, and that they can offer a scalable alternative to conventional domain-of-stability estimation methods in islanded distribution systems. The implementation of cGANs-based stability assessment is described for an LV distribution test case and its advantages demonstrated.
\end{abstract}


\IEEEpeerreviewmaketitle

\section{Introduction}

In distribution networks that are tied to weak grids or islanded, decentralized power sharing mechanisms are implemented for forming the grid. The most common control strategy is the $P$-$f$/$Q$-$V$ droop control wherein the voltage and frequency of each source varies as per a linear law based on the real and reactive power output of that source. Such a control strategy could manifest poorly damped poles for some values of the droop parameters \cite{chang2014dynamical}, and the distribution system operator's fast response to these poorly-damped power flows is imperative to the continuing operation of the grid. 

The small-signal stability of the system is dependent on the network topology, loading, generation and the power-sharing controller parameters. Before the system begins operation, offline tuning is performed to obtain the optimal values of the droop coefficients while considering constraints such as the maximum steady-state voltage and frequency deviations, and the desired power outputs of each source. However, the generation levels can change frequently at the distribution level with the presence of highly variable renewable generation \cite{lee2013power}. Moreover, the network configuration could also be significantly affected by tap-changes, line switching, and faults. To assess the stability of such grids in real time, their eigenvalues must be examined based on the real-time conditions.

To maintain real-time stability, the first approach is to use the nominal system configuration to design the droop values in an offline manner while allowing a sufficient margin from the instability limit. The expectation here is that the system will remain within the stable region as the operating point changes. However, such a conservative droop selection will lead to a poor power-sharing performance when there is significant deviation from the assumed operating point \cite{Simpson-Porco2015}. An alternative approach proposed in \cite{Zhang2016} effects real-time corrections on the droop coefficients to achieve less conservative settings using a global stability indicator. While this approach is more advantageous than offline tuning, it does not actually determine the domain-of-stability (i.e., the hyperspace of all droop gains that yield stable behavior) in real-time, and therefore still yields somewhat conservative results. If the full domain-of-stability  were to be known accurately, appropriate droop selection can be made with the required stability margin.

Conventionally, the determination of the stability region entails the evaluation of eigenvalues for a range of droop gains, and the stability classification of each setting. This process has a complexity $O(n^3)$ based on the number of states $n$ (detailed models contain 5 states per source) \cite{Nikolakakos2015}. Consequently, the stability region determination could take several tens of seconds or minutes for large systems during which time low-inertia systems could well face tripping of lines/sources. An alternative approach would be to develop an \textit{a priori} database of various possible system configurations with the domain-of-stability for each case. While this may address the computation time issue with online stability assessment, it raises other concerns pertaining to the development of an exhaustive database, its non-adaptability to changes in configuration, and the time required to identify the relevant database entry for a given real-time configuration.

Deep learning techniques have been applied for studying multi-inverter dynamics in order to obtain fast black-box models (most recently in \cite{Amoateng2018}). However, their use has been largely limited to learning time-domain behavior. In contrast, the focus of this work is on the frequency-domain behavior. This paper proposes the use of Generative Adversarial Networks (GANs), specifically, conditional GANs (cGANs) for obtaining the stability hyperspace of control parameters. This study is a novel demonstration of the ability of cGANs to learn the notion of small-signal stability, and generate the complete stability region in real-time for the present distribution network configuration. As a result, this guarantees stable operation while enabling non-conservative droop settings to be selected so as to achieve an optimal power-sharing performance. The cGANs are trained offline, and while online, the computational time for generating the stability region is demonstrated to be significantly lower than that for the traditional tuning method. It will also be demonstrated that the training process itself is scalable to the number of system configurations, implying that the proposed cGANs approach will be comparable, or better than the look-up-table approach as well.

\section{Conventional Stability Region Determination}

The system of interest here is a distribution network with several conventional and power electronic sources, each governed by the following droop equations: 
\begin{eqnarray}\nonumber
f=f_0-k_f \left[\frac{\omega_c}{s+\omega_c}\right] (P-P_0) \text{, and} \\
V=V_0-k_v \left[\frac{\omega_c}{s+\omega_c}\right] (Q-Q_0),
\label{droop}
\end{eqnarray}
where $f$, $V$, $P$, and $Q$ respectively denote the frequency, terminal voltage magnitude, real and reactive power injections of each source. The subscript `0' indicates their respective nominal values. $k_f$ and $k_v$ are the $P$-$f$ and $Q$-$V$ droop coefficients respectively, and $\omega_c$ is the first-order power filter corner frequency. While the conventional droop is taken up here for proof-of-concept, the proposed method is equally applicable to more sophisticated droop control strategies such as opposite droop \cite{chang2014dynamical}.

The eigenvalues of the system are obtained as the solution of:
\begin{eqnarray}
[\mathbf{A} +  \mathbf{B} s +  \mathbf{C} s^2 +  \mathbf{D} s^3 +  \mathbf{E} s^4]
\begin{bmatrix} \Delta \theta \\ \Delta V \end{bmatrix} = 0.
\label{eveqn}
\end{eqnarray}
The coefficient matrices are defined as follows:
\[
A=
\begin{bmatrix}
-(\rho^2+1) \mathbf{B}  &  -\rho (\rho^2+1) \mathbf{B} \\
\rho (\rho^2+1) \mathbf{B}  &  (\rho^2+1) (-\mathbf{B}+\mathbf{L_q})
\end{bmatrix}
\]
\[
B=
\begin{bmatrix}
(\rho^2+1) \mathbf{L_p}  &  -(\rho^2+1) \frac{\mathbf{B}}{\omega_0} \\
(\rho^2+1) \frac{\mathbf{B}}{\omega_0}  &  ((\rho^2+1) T + 2 \frac{\rho}{\omega_0}) \mathbf{L_q}
\end{bmatrix}
\]
\[
C=
\begin{bmatrix}
((\rho^2+1) T + 2 \frac{\rho}{\omega_0}) \mathbf{L_p}  &  \mathbf{0} \\
\mathbf{0}  &  (\frac{1}{\omega_0^2} + 2 \frac{\rho T}{\omega_0}) \mathbf{L_q}
\end{bmatrix}
\]
\[
D=
\begin{bmatrix}
(\frac{1}{\omega_0^2} + 2 \frac{\rho T}{\omega_0}) \mathbf{L_p}  &  \mathbf{0} \\
\mathbf{0}  &  \frac{T}{\omega_0^2} \mathbf{L_q}
\end{bmatrix}
\]
\[
E=
\begin{bmatrix}
\frac{T}{\omega_0^2} \mathbf{L_p}  &  \mathbf{0} \\
\mathbf{0}  &  \mathbf{0}
\end{bmatrix}
\]
where $\mathbf{G}+j\mathbf{B}=\mathbf{Y_\mathit{bus}}$ is the bus-admittance matrix of the distribution network, $\mathbf{L_p}$ and $\mathbf{L_q}$ diagonal matrices containing the inverse of $k_f$ and $k_v$ respectively of all the sources, $\rho$ the $R/X$ ratio of the system, $\omega_0$ the nominal power frequency ($100\pi$ rad/s), and $T=1/\omega_c$. To determine the stability region numerically, a hyperspace of possible droop coefficients is first conceptualized, and the stability of each point is determined to obtain the full domain-of-stability. A droop setting is considered stable if the real parts of all eigenvalues are negative.

\section{Stability Region Determination using cGANs} 

\subsection{Review of GANs and cGANs}

GANs have been used in the image processing domain to create synthetic images from a training set of real images, for example, pictures of human faces, birds,  etc. GANs consist of two neural networks- a Generator and a Discriminator. The former generates new data sets, and the latter judges how similar the generated data set is to the training data (real data); the two networks compete to improve their respective accuracies during training \cite{goodfellow2014generative}. The following is a mathematical overview of the GANs training process.

Let $x$ be the actual data with a distribution $p_\mathit{data}$. The Generator is fed noise $z$, which maps it to $G(z)$. The Discriminator, when fed an input $x$, produces a single scalar $D(x)$ that represents the probability that $x$ came from the training data rather than from the Generator. The output of the Generator is fed to the Discriminator, and the networks are trained simultaneously. During the course of the training, the goal is to maximize the accuracy of the Discriminator, while minimizing $\text{log}(1-D(G(z)))$. This is equivalent to the minimax game with a value function $V(D,G)$ given by:
\begin{eqnarray}\nonumber
\underset{G}{\text{min}}\, \underset{D}{\text{max}}\text{V}\left(D,G\right)=
\mathbb{E}_{x\thicksim p_\mathit{data}(x)}[log D(x)]+ \\ \mathbb{E}_{z\thicksim p_{z}(z)}[log(1-D(G(z)))].
\label{gan}
\end{eqnarray}
At the end of the training process, the distribution of the generated (synthetic) data $p_g$ becomes equal to $p_\mathit{data}$. The Discriminator is therefore unable to differentiate between the real data and the generated data, i.e., $D(x)=D(G(z))=0.5$.

Conditional GANs (cGANs) entail an additional input $y$ to both $G$ and $D$, which can be used to impose a condition on the generated data samples \cite{mirza2014conditional}. For example, cGANs can generate pictures of \textit{smiling} faces from a training dataset of faces with several expressions. The training equation in this case is:
\begin{eqnarray}\nonumber
\underset{G}{\text{min}}\, \underset{D}{\text{max}}\text{V}\left(D,G\right)=
\mathbb{E}_{x\thicksim p_{data}(x)}[log D(x\rVert y)]+ \\
\mathbb{E}_{z\thicksim p_{z}(z)}[log(1-D(G(z\rVert y)))].
\label{cgan}
\end{eqnarray}

\subsection{Application to distribution system stability studies}

Since GANs in general have the capability to generate new data sets with similar characteristics as the training data, they can be usefully leveraged for power system stability characterization. The principle behind using cGANs instead of simple GANs is the following. The data representing each power system includes the network parameters and droop coefficients. When trained with several possible stable system configurations and droop settings, simple GANs will generate additional stable cases, thus generating a stability hyperspace (droop coefficients) for a mixture of network configurations. When cGANs are used with the conditional input as the present network configuration, they will directly provide the stability region for that particular configuration. Notably, since the training of the cGANs will be performed offline, the real-time generation of the domain-of-stability can be obtained relatively faster when compared to the conventional method, as will be demonstrated in the following subsections.

\subsection{Implementation details}

\begin{figure}
	\centering
	\includegraphics[width=2.2in]{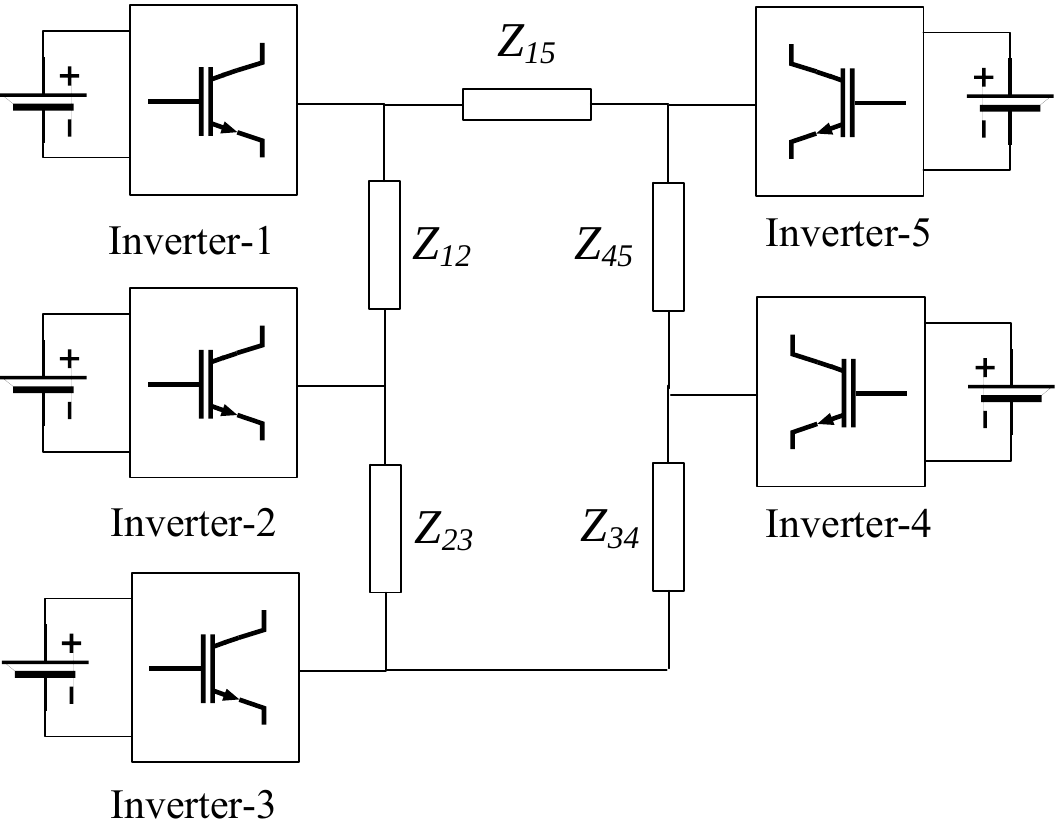}
	\vspace*{-2mm}
	\caption{Schematic of 5-node ring-main distribution system.}
	\vspace*{-2mm}
	\label{Topology}
	\vspace*{-2mm}
\end{figure}

\begin{table}
	\renewcommand\arraystretch{1.3}
	\centering
	\caption{Network Parameters of Test System}
	\vspace*{-2mm}
	\begin{tabular}{|c|c|c|c|}
		\hline			
		Impedance & Value ($\Omega$) & Impedance & Value ($\Omega$) \\
		\hline
		$Z_\mathit{12}$ & $0.08+j0.08$ & $Z_\mathit{45}$ & $0.15+j0.15$\\
		$Z_\mathit{23}$ & $0.15+j0.15$ & $Z_\mathit{15}$ & $0.02+j0.02$\\
		$Z_\mathit{34}$ & $0.05+j0.05$ & & \\	
		\hline
	\end{tabular}%
	\label{5busparam}%
	\vspace*{-4mm}
\end{table}

Online stability region determination using cGANs is demonstrated using a 4.16kV, 50Hz ring system shown in Fig. \ref{Topology}. The power rating of each of the five identical sources is 1 MVA, and the nominal droop coefficients are $k_f$=0.15\% and $k_v$=5.0\%. 
The network impedances are presented in Table \ref{5busparam}. The power filter cut-off frequency $\omega_c$ is taken as 31.41 rad/s. Let $\mathbf{Y_\mathit{bus}=Y\angle} \boldsymbol{\theta}$ be the bus admittance matrix of the network, and $\mathbf{k_f}$ and $\mathbf{k_v}$ be vectors including the droop coefficients of all the sources. The training dataset is created in MATLAB, with each data point consisting of $\mathbf{Y}$, $\boldsymbol{\theta}$, $\mathbf{k_f}$, and $\mathbf{k_v}$. Together, these four parameters determine the stability of the system. The elements of the above matrices/vectors are concatenated to obtain a single vector; positional information is not recognized by non-convolutional networks. While generating the data, as these matrices/vectors have elements having different ranges of magnitudes, these are scaled by dividing with the largest element of the respective matrices, bringing their magnitudes in the interval $[0,1]$. This scaling factor is uniformly used for all the data points (i.e., different power system configurations), guaranteeing that due importance is given to the quantities with smaller magnitudes during the training, while accelerating it \cite{simard1998transformation}. The inverse is multiplied to regain the true physical parameters from the cGANs output.

The cGANs are realized using the Pytorch package in Python and executed on a PC running 64-bit Windows-10 OS, with an i7-8550 processor and 8GB RAM. The Generator and Discriminator entail a fully connected neural network with 4 and 3 layers respectively. The activation function for all layers is LeakyRelu, except the Discriminator's output layer, which is the sigmoid function. For each epoch, small-batch training is done and for the Generator, batch normalization is performed so as to accelerate the optimization process. This reduces the algorithm's sensitivity to the learning rate \cite{ioffe2015batch}.

To guage the performance of the Discriminator during the training process, we define the term ``real loss" to denote the cross-entropy between $D(\mathbf{x})$ and the unit vector, where $\mathbf{x}$ is the training data set. Similarly, ``fake loss" is the cross-entropy between $D(G(\mathbf{z}))$ and the zero vector, where $\mathbf{z}$ is a random vector. For the Generator, ``G loss" is the cross-entropy between $D(G(\mathbf{z}))$ and the unit vector. When the training process is completed, all of these three quantities approach log(2)=0.69 because $D(G(\mathbf{z}))$ and $D(\mathbf{x})$ both tend to 0.5 as mentioned in Section III A.

The output data points of the cGANs correspond to synthetic power system configurations and their parameters are expected to conform to practical values. Thus, the follow stopping criterion is used for the training process:
\begin{eqnarray}
d_c = \underset{z}{\text{max}} (x_f-G(z)_f) < \epsilon,
\label{stop_crit}
\end{eqnarray}
where $d_c$ is the Chebyshev distance, and $\epsilon$ is the allowable deviation of the generated data from the real data. As the data matrices $\mathbf{Y}$, $\boldsymbol{\theta}$, $\mathbf{k_f}$, and $\mathbf{k_v}$ have already been scaled to $[0,1]$, $\epsilon$ can be uniformly be used for all, and this value is selected as 0.05 for this work. 

\subsection{Simple GANs for a single network configuration}

\begin{figure}
	\centering
	\subfigure[Real\,loss]{\includegraphics[width=2.5in]{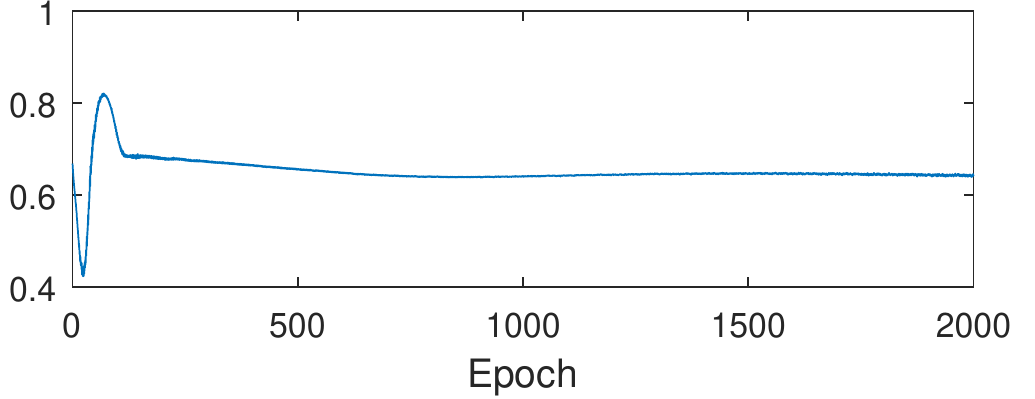}}\\[-1mm]
	\subfigure[Fake\,loss]{\includegraphics[width=2.5in]{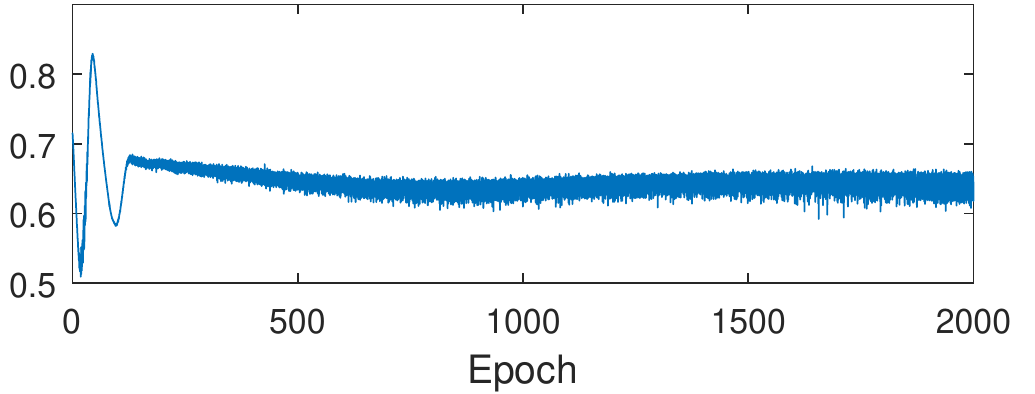}}\\[-2mm]
	\subfigure[G\,loss]{\includegraphics[width=2.5in]{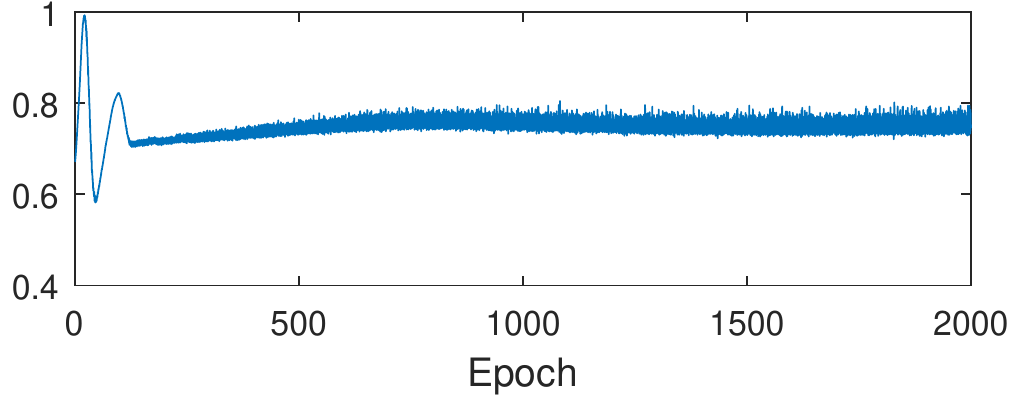}}\\[-2mm]
	\subfigure[Chebyshev\,distance]{\includegraphics[width=2.5in]{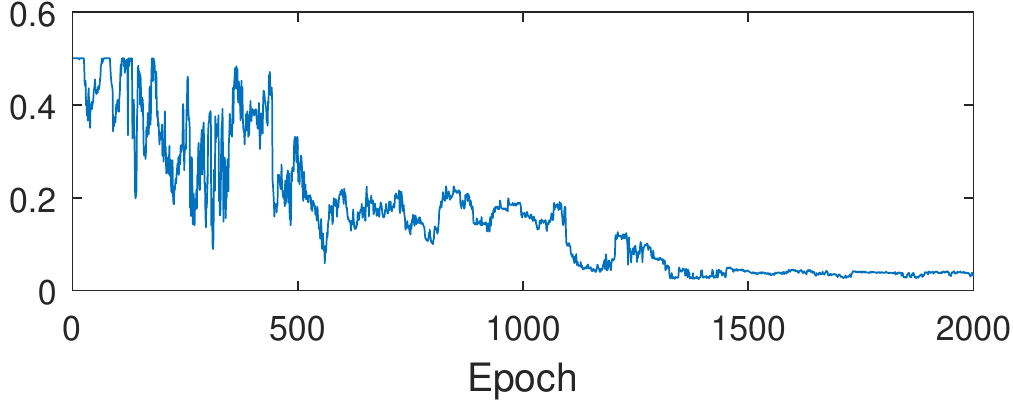}}\\[-2mm]	
	\caption{Training procession for simple GANs with learning rate $8\times10^{-6}$.}
	\label{res_single_detailed}	
	\vspace*{-2mm}
\end{figure}

First, we consider simple GANs to demonstrate its usefulness for this application, and highlight the need to use cGANs, which is employed in the following subsection. To obtain the domain of stability with respect to the first inverter (for ease of representation on a 2-D plane), the values of $k_\mathit{fi}$ and $k_\mathit{vi}$ for $i$=2-5 are respectively fixed at 0.1\% and 2\%, and the training data set is generated by varying $k_\mathit{f1}$ and $k_\mathit{v1}$ uniformly in the range $[0,0.4]$ and $[0,4]$ respectively. The training set size is chosen as 4000, the batch size 100, and learning rate $8\times 10^{-6}$. The real loss, fake loss, G loss and $d_c$ for each epoch are recorded as shown in Fig. \ref{res_single_detailed}. 

\begin{figure}[t]
	\centering
	\subfigure[Epoch 1130 (Accuracy=99.38\%)]{\includegraphics[width=1.7in]{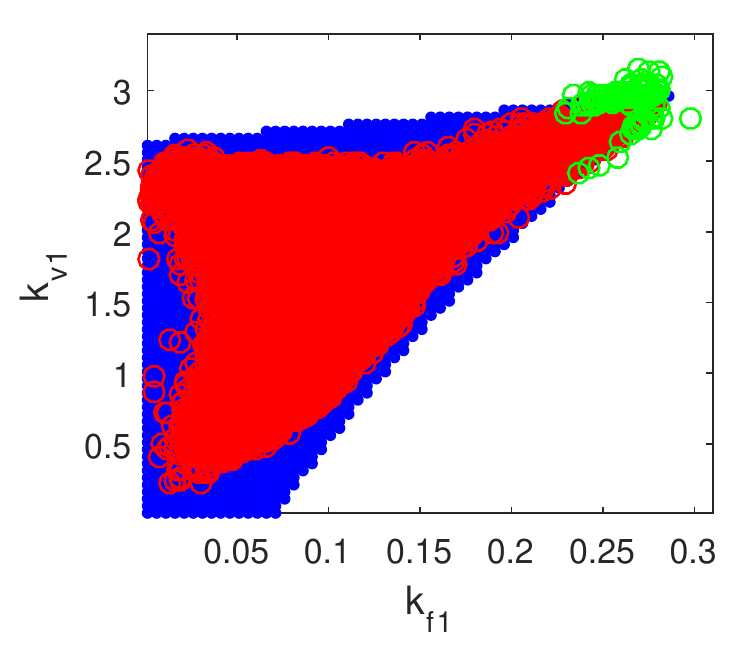}}
	\subfigure[Epoch 1900 (Accuracy=98.95\%)]{\includegraphics[width=1.7in]{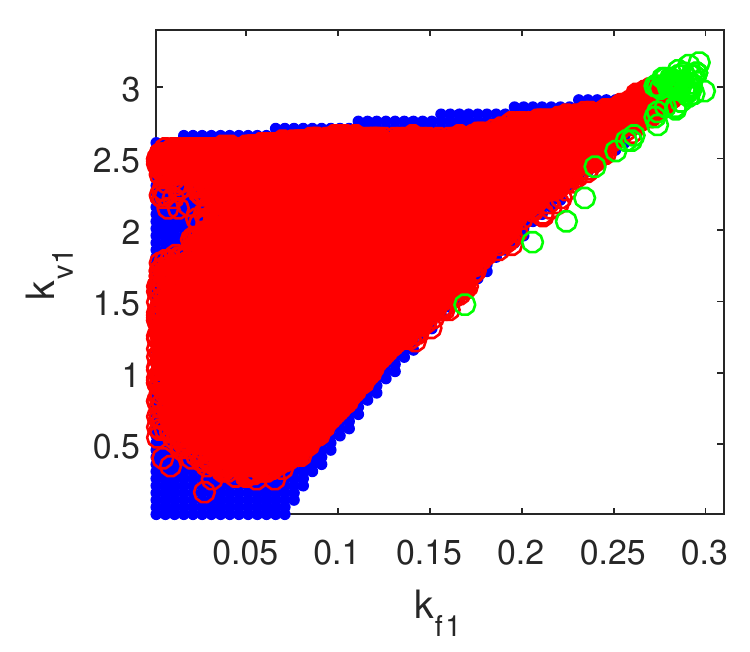}}\\[-2mm]
	\caption{Stability region obtained from 20000 samples from GANs is plotted in red. The theoretical stability region is shown in blue. The points identified by the GANs not in the actual stability region are shown in green.}
	\label{res_single}
	\vspace*{-6mm}
\end{figure}

Although the loss functions reach their final values around epoch 500, the goal (\ref{stop_crit}) is first achieved at epoch 1130. The value $d_c$, when smaller, indicates higher generation accuracy, and that a wider region of the actual stability region will be populated. Random noise is then fed to the trained GANs to populate the stability region shown in Fig. \ref{res_single}(a) with 20000 samples. The theoretical stability region obtained from the traditional numerical method is also shown. The data samples from the GANs are found to be 99.38\% accurate. With further training, the GANs generates a better coverage of the stability region, as seen from Fig. \ref{res_single}(b) corresponding to epoch 1900. 

Some of the identified points are marked to be actually ``unstable" even though they are just inside the boundary of the stability region. These correspond to GANs output samples that have a slight variation in the $\mathbf{Y}$ and $\boldsymbol{\theta}$ which makes them actually unstable; such variations are inherent to GANs and therefore the optimal droop setting should be picked sufficiently farther from the boundaries to guarantee stability. However, this error can be further reduced if desired, through additional epochs of training.

The need for cGANs is clear from this case because, if a variety of system topologies were used to generate the training data, then the output would be spread out among those topologies as well, generating a large set of stable cases. However, since the application calls for selecting the appropriate droop settings for the present system configuration, cGANs can be leveraged, with the conditional label being the real-time $\mathbf{Y}$ matrix.   

\subsection{Conditional GANs for multiple network configurations}

\begin{figure}
	\centering	
	\subfigure[Real\,loss]{\includegraphics[width=2.5in]{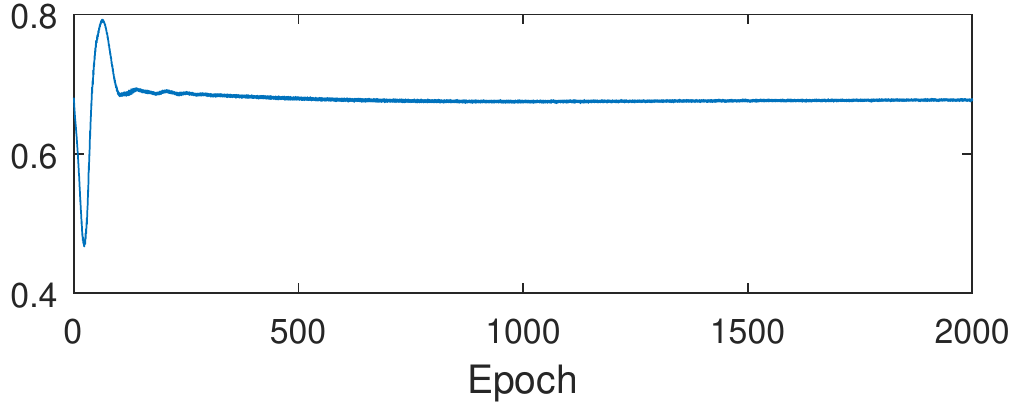}}\\[-1mm]
	\subfigure[Fake\,loss]{\includegraphics[width=2.5in]{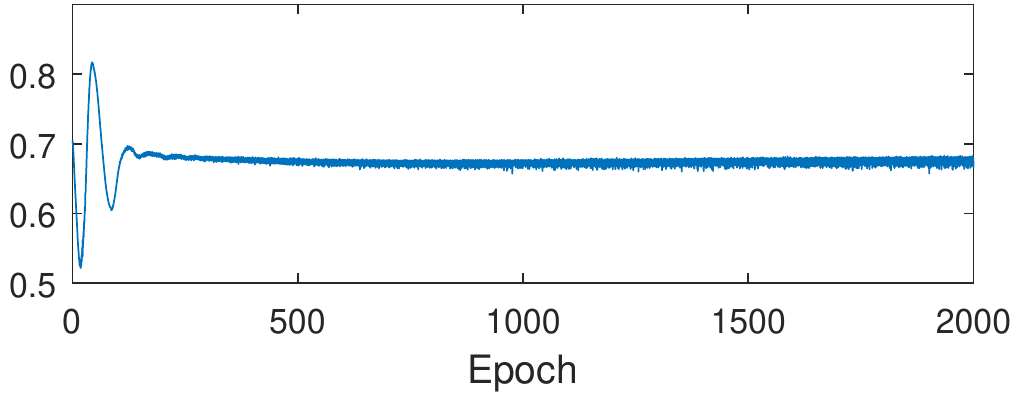}}\\[-2mm]
	\subfigure[G\,loss]{\includegraphics[width=2.5in]{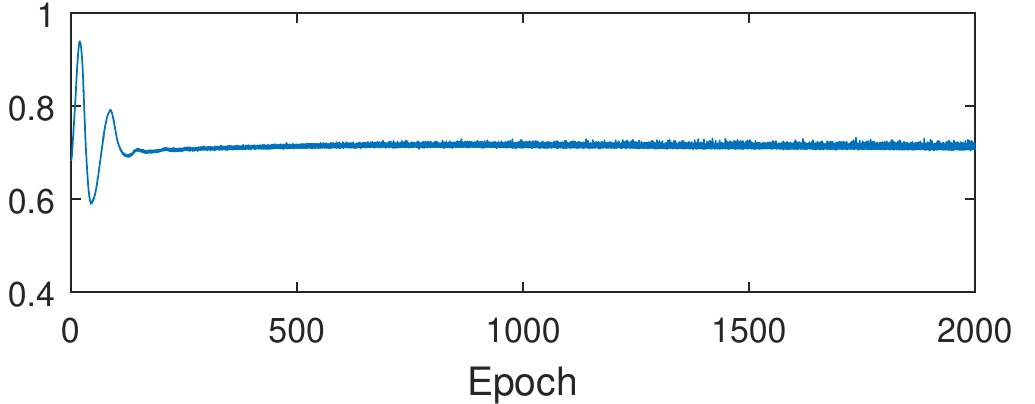}}\\[-2mm]
	\subfigure[Chebyshev\,distance]{\includegraphics[width=2.5in]{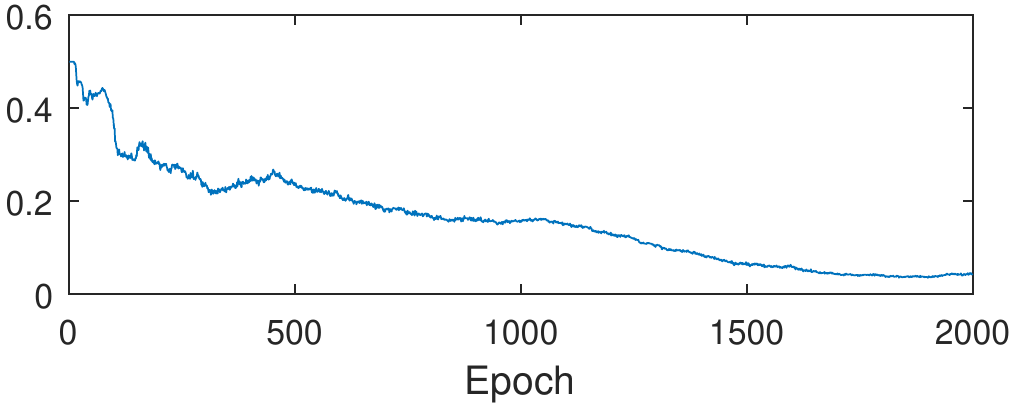}}\\[-2mm]
	\caption{Training procession for cGANs with learning rate $8\times10^{-6}$. }
	\label{res_cgan_detailed}
	\vspace*{-6mm}
\end{figure}

\begin{figure}
	\centering
	\subfigure[Original (Accuracy=98.43\%)]{\includegraphics[width=1.7in]{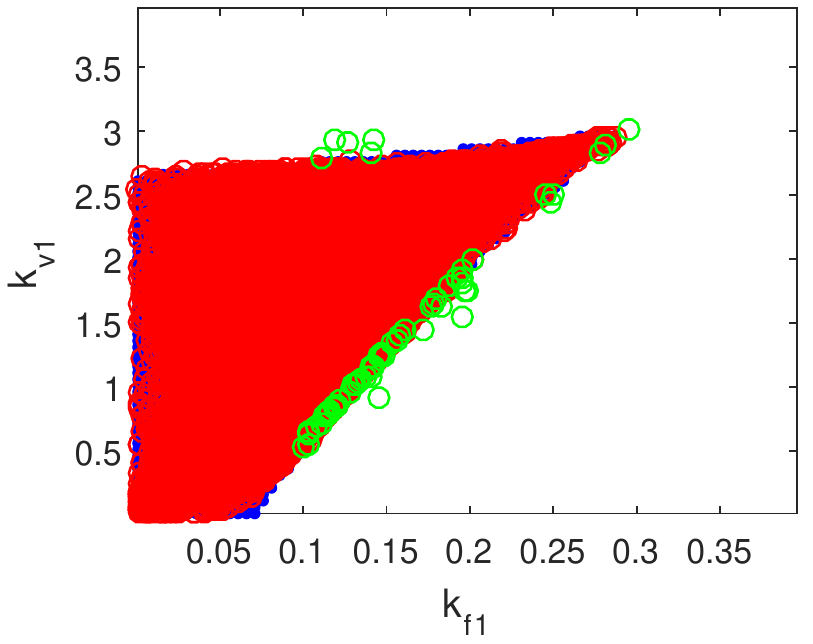}}
	\subfigure[$y_{12}/2$ (Accuracy=98.78\%)]{\includegraphics[width=1.7in]{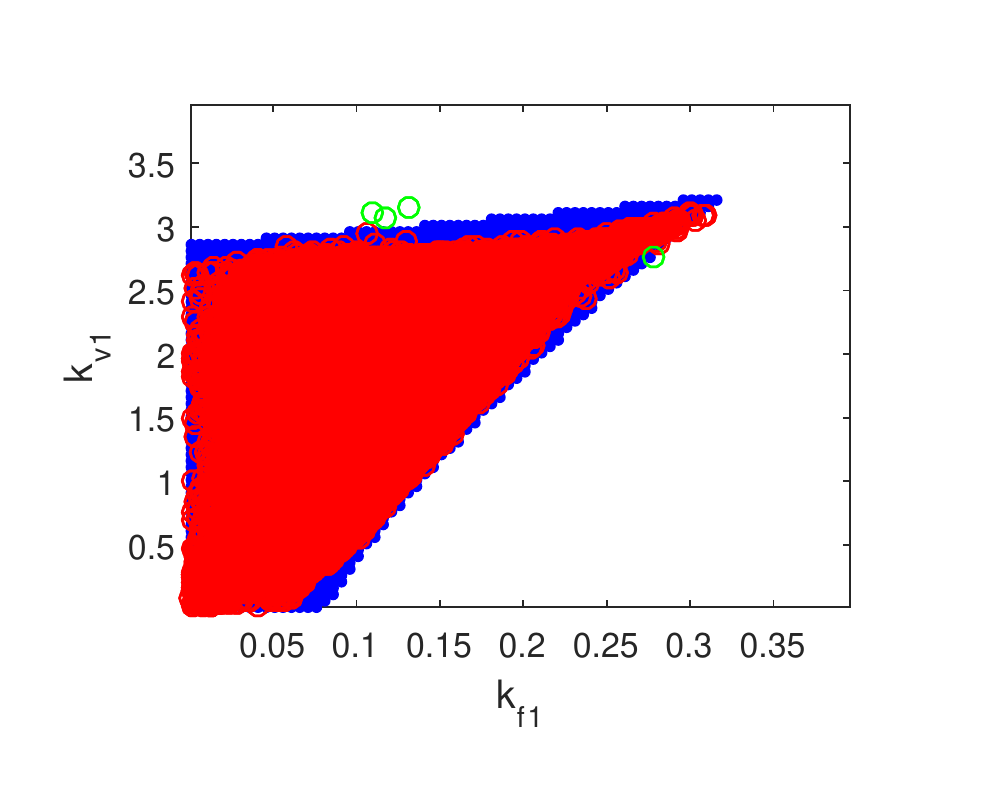}}\\[-1mm]
	\subfigure[$y_{23}/2$ (Accuracy=97.85\%)]{\includegraphics[width=1.7in]{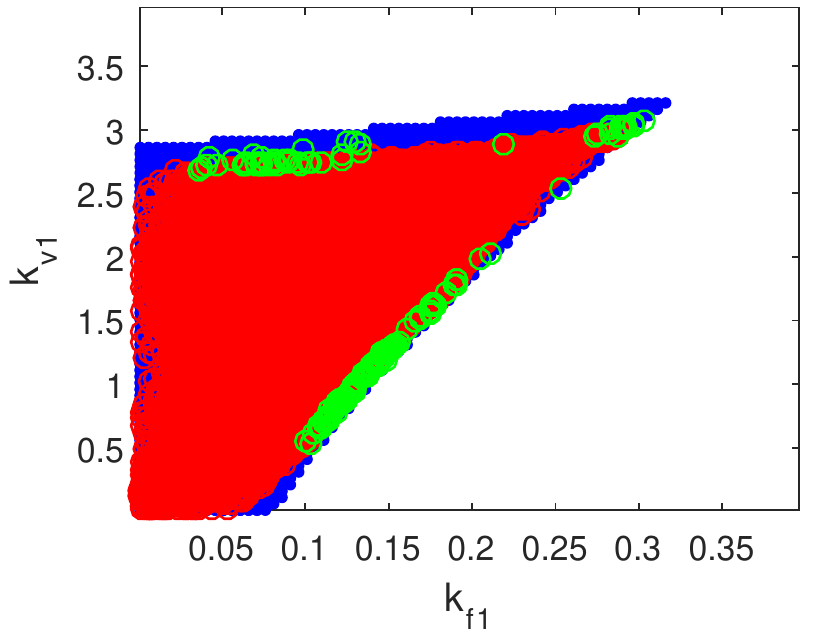}}
	\subfigure[$y_{34}/2$ (Accuracy=99.80\%)]{\includegraphics[width=1.7in]{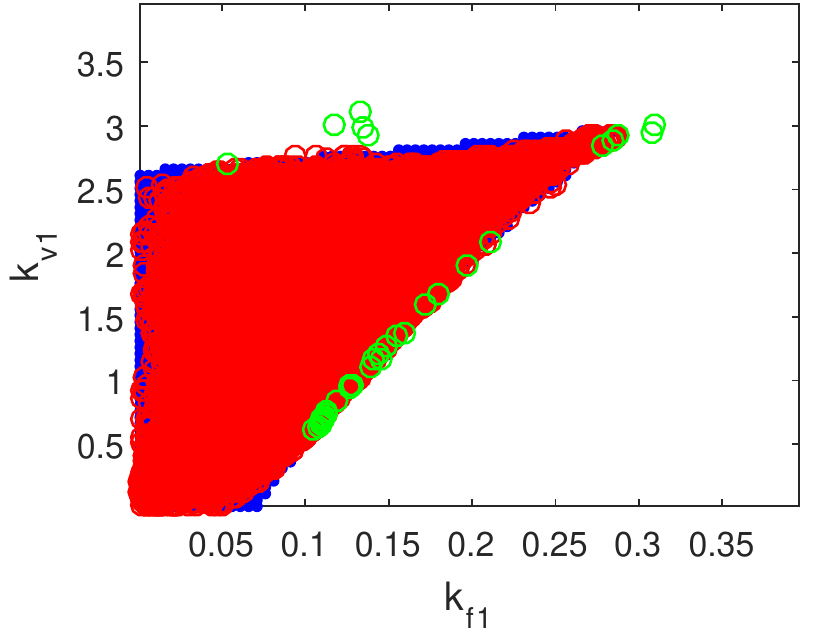}}\\[-1mm]
	\caption{Stability region (shown in red) for Inverter-1 identified using cGANs for 4 system configurations at epoch 1907. Corresponding theoretical regions are indicated in blue, obtained from the traditional method. Green circles denote erroneously projected points of stability by cGANs method.}
	\label{res_cgan}
	\vspace*{-3mm}
\end{figure}

The use of cGANs is demonstrated considering the same system as before, along with certain contingencies. Assuming that there are two parallel feeders between Nodes 1-2, 2-3 and 3-4, we consider the cases of loss of one of these parallel connections separately (i.e., $y_{ij}$ is halved), yielding 3 additional distribution system configurations apart from the original system. The structure of each training data vector is the same, covering 16000 samples equally distributed over the 4 system configurations. The batch size is set as 400 and $d_c$ is calculated every 20 epochs for each of the 4 configurations by providing the appropriate conditional label to the cGANs. The maximum $d_c$ of all the configurations is considered to be the $d_c$ for the generated data. The criterion (\ref{stop_crit}) is satisfied at epoch 1630, but in the interest of better populating the stability region, the training is carried out for over 300 additional epochs, and the relevant plots are shown in Fig. \ref{res_cgan_detailed}. The corresponding cGANs are used to generate 20000 samples for each of the 4 system configurations to obtain their respective stability regions, which are shown in Fig. \ref{res_cgan}. It is clear that this is different for each system as expected, and that the accuracy of the output samples is high, above $97.5\%$ for each case.

To demonstrate the computational advantage of the cGANs vis-à-vis the traditional approach, both models are used to generate 20000 sets of stable $(k_\mathit{f_1}, k_\mathit{v_1})$ samples for each system configuration. The time for the sample generation is noted in Table \ref{time_comp}. First, the droop coefficients for Inverters 2-5 are kept constant, i.e., the stable hyperplane corresponding to the first inverter alone is generated. Second, as is practically necessary, the sample generation is carried out varying the droop parameters of all the inverters to generate the whole stability hyperspace, and the corresponding runtime is also noted in the same table. For this, the learning rate is selected as $2 \times 10^{-5}$ and the selections of $k_\mathit{v_i}$ and $k_\mathit{f_i}$ ($i$=1-5) obey uniform distribution in their respective ranges of $[1,5]$ and $[0.1,0.5]$. It is observed that for any particular system configuration, the traditional domain-of-stability determination approach takes well above 10s, but the cGANs runs for just over 1.2s, with good accuracy. This indicates the suitability of cGANs for practical deployment.

\begin{table}
	\renewcommand\arraystretch{1.3}
	\centering
	\setlength{\tabcolsep}{4pt}
	\caption{Running Time for 20000 Samples- Accuracy in Parentheses}
	\vspace*{-1mm}
	\begin{tabular}{|c|c|c|c|c|c|}
		\hline
		\multicolumn{1}{|p{0.1cm}|}{Approach} & \multicolumn{1}{c|}{$k_\mathit{f_i} \& k_\mathit{v_i}$} & \multicolumn{1}{c|}{System a} & \multicolumn{1}{c|}{System b} & \multicolumn{1}{c|}{System c} & \multicolumn{1}{c|}{System d}\\		
		&($i$=2-5) & Time (s) & Time (s) & Time (s) & Time (s)\\
		\hline
		Traditional & Fixed & 14.2336 & 16.5988 & 15.1759 & 14.8897\\		 
		\hline
		Traditional & Varied &  12.1833 & 12.3569 & 12.4272 & 12.4838\\	
		\hline
		cGANs & Fixed & 1.3480 & 1.3776 & 1.3343 & 1.2241\\
		Epoch 1907 &&($98.43\%$)&($98.78\%$)&($97.85\%$)&($98.03\%$)\\
		\hline
		cGANs & Varied & 1.2838 & 1.2896 & 1.3115 & 1.2256\\
		Epoch 3994 &&($100\%$)&($99.98\%$)&($99.97\%$)&($99.80\%$)\\	 
		\hline			
	\end{tabular}%
	\label{time_comp}%
	\vspace*{-4mm}
\end{table}%

\begin{table}[!]
	\renewcommand\arraystretch{1.3}
	\centering
	\caption{Required Epochs For Learning Rate $8 \times 10^{-6}$}
	\vspace*{-2mm}
	\begin{tabular}{|l|c|c|c|}
		\hline			
		Number of system configurations & 1 & 2 & 4 \\
		\hline
		Epoch number with accuracy above 95\% & 680 & 695 & 720\\
		Epoch number with whole region populated & 1035 & 1200 & 1280\\
		\hline
	\end{tabular}%
	\label{epoch_table}%
	\vspace*{-5mm}
\end{table}%

\subsection{Demonstration of scalability}

The scalability of cGANs in determining the domain-of-stability stems from the fact that it can generate the plots corresponding to multiple system configurations without additional real-time computational complexity as compared to the simple GANs. However, the training process demands more data. Further, as the number of system configurations in the training dataset increases, the number of epochs required for good accuracy, as well as for populating the entire stability region increases as evident from Table \ref{epoch_table}. However, the required number of epochs does not increase significantly when the number of system configurations increases, demonstrating the scalability of the training process. Therefore, the cGANs approach is expected to perform better, and in the worst case, equal, to a look-up-table approach.

\section{Conclusion}

This study has demonstrated that the computational time requirements for online domain-of-stability characterization of networked droop controlled sources can be met by cGANs. The cGANs-based method can accurately yield the stability region for the present system configuration, which can be effectively used by the supervisory controller to make real-time adjustments to the droop controllers for ensuring small-signal stability. Importantly, these  droop settings can be selected with a flexible degree of conservatism. The scalability of the proposed method with respect to the required training epochs and number of possible system configurations has been demonstrated. Comparison of the running time between the cGANs-based and conventional methods indicate that the former is nearly 10 times faster in generating the stability region for each system configuration, indicating its effectiveness for supervisory control. Future work should address the optimal selection of the training parameters such as learning rate, and the size of the training dataset so as to achieve high accuracy and stability region coverage. 

\section*{Acknowledgment}

This work was supported by the Singapore Ministry of Education grant R-263-000-D10-114.

\balance

\bibliographystyle{IEEEtran}  
\bibliography{ref}

\end{document}